\documentclass[prl,superscriptaddress,twocolumn,showpacs]{revtex4}

\usepackage{enumerate}
\usepackage{amsfonts,amssymb,amsmath}
\usepackage[]{graphics,graphicx,epsfig}
\usepackage{amsthm}

\usepackage{graphicx}
\usepackage{dcolumn}
\usepackage{natbib}
\usepackage{color}
\usepackage{multirow}

\def\identity{\leavevmode\hbox{\small1\kern-3.8pt\normalsize1}}

\newtheorem{propo}{Proposition}
\newcommand{\be}{\begin{eqnarray}}
\newcommand{\ee}{\end{eqnarray}}
\newcommand{\bpr}{\begin{propo}}
\newcommand{\epr}{\end{propo}}
\newcommand{\bpf}{\begin{proof}}
\newcommand{\epf}{\end{proof}}

\renewcommand{\epsilon}{\varepsilon}

\begin{document}


\title{Generalized Probabilistic Description of Noninteracting Identical Particles}

\author{Marcin Karczewski}   
\affiliation{Faculty of Physics, Adam Mickiewicz University, Umultowska 85, 61-614 Pozna\'n, Poland}

\author{Marcin Markiewicz}     
\affiliation{Institute of Physics, Jagiellonian University, ul. \L{}ojasiewicza 11, 30-348 Krak\'ow, Poland}

\author{Dagomir Kaszlikowski}
\affiliation{Centre for Quantum Technologies,
National University of Singapore, 3 Science Drive 2, 117543 Singapore,
Singapore}
\affiliation{Department of Physics,
National University of Singapore, 3 Science Drive 2, 117543 Singapore,
Singapore}

\author{Pawe\l{} Kurzy\'nski}   \email{pawel.kurzynski@amu.edu.pl}   
\affiliation{Faculty of Physics, Adam Mickiewicz University, Umultowska 85, 61-614 Pozna\'n, Poland}
\affiliation{Centre for Quantum Technologies, National University of Singapore, 3 Science Drive 2, 117543 Singapore, Singapore}

\date{\today}


\begin{abstract}
We investigate an operational description of  identical noninteracting particles in multiports. In particular we look for physically motivated restrictions that explain their bunching probabilities. We focus on a symmetric 3-port in which a triple of superquantum  particles admitted by our generalized probabilistic framework would bunch with probability $\frac{3}{4}$. The bosonic bound of $\frac{2}{3}$ can then be restored by imposing the additional requirement of product evolution of certain input states. These states are characterized by the fact that, much like composite systems, their entropy equals the sum of entropies of its one-particle substates. This principle is however not enough to exclude the possibility of superquantum particles in higher-order multiports.

\end{abstract}

\maketitle


\emph{Introduction.--}Although quantum mechanics is a well established theory, its foundations still lack a satisfactory explanation. This is in a stark contrast with the special relativity, where all the predictions can be traced back to the invariance of the physical laws in inertial systems and the constant speed of light for all observers. To deepen our understanding of the quantum theory, we need to discover the underlying principles.

This quest has been undertaken in two ways. First of them, the \emph{device-independent} approach, consists in restricting the conditional probability distributions of some black boxes with information-theoretic principles. For example, in their seminal paper 
 \cite{Popescu1994} Popescu and Rohrlich proposed a principle, which guarantees that a box cannot be used for superluminal communication. This restriction, called no-signaling, was however not enough to exclude all stronger than quantum correlations. Soon more fundamental principles were discovered, including macroscopic locality \cite{navascues2009glance}, local orthogonality \cite{fritz2013local} and many others \cite{brassard2006limit,sinha2010ruling,pawlowski2009information}, but none of them was fully successful in restoring the quantum theory.
 
The other approach, known under the umbrella term \emph{generalized probabilistic theory} (GPT), aims to single out the quantum formalism from  information-theoretic principles. It defines the notions of systems, states, transformations and measurements and then narrows them down with additional axioms until Hilbert spaces, density matrices and the Born rule appear. Some  notable works written in this spirit include \cite{chiribella2011informational,de2012deriving,masanes2011derivation,hardy2001quantum,dakic2009quantum}.

In this work we employ elements of the GPT formalism to provide an operational description of  linear optical interferometric experiments with bosons and fermions. Our framework consists in input and output probability distributions of states linked with a transformation matrix. This matrix captures two important features of optical multiports. Firstly, the  particles do not interact, therefore each particle evolves individually and the differences in measured probability distributions stem solely from the particle-statistics and interference \cite{PhysRevLett.59.2044,pittman1996can,kim2003two,PhysRevA.62.013809,PhysRevLett.111.130503,tichy2012many,tichy2014interference,omar2005indistinguishable}. To reflect that we impose a consistency condition which constrains transformations of probability distributions. This condition is analogous to no-signaling and states that the distribution of an individual particle, or a subset of particles, cannot depend on the total number of particles. 
Secondly, the information is not erased, which is implemented by requiring that the transformation matrix be doubly stochastic \cite{cover}.

These restrictions are obeyed by quantum particles, however we show that they allow for existence of hypothetical particles whose grouping tendencies, commonly known as bunching, are stronger than in case of bosons. This example can be considered as an analog of a Popescu-Rohrlich (PR) box within the realm of identical particles. Finally, we provide an additional principle which rules out the superbunching particles on a tritter (symmetric 3-port). It consists in requiring that states whose entropy  equals the sum of the entropies of its substates  undergo a product evolution. Such states resemble composite systems, which are significant components of many GPTs (see for instance \cite{hardy2001quantum,dakic2009quantum,masanes2011derivation}). Interestingly, this principle is not enough to exclude superquantum particles in higher-order multiports.

The motivation for our research is twofold. Firstly, we would like to contribute to the search for general rules underlying the foundations of quantum theory \cite{Popescu1994,navascues2009glance,fritz2013local,brassard2006limit,sinha2010ruling,pawlowski2009information,chiribella2011informational,
masanes2011derivation,de2012deriving,hardy2001quantum,dakic2009quantum}. In particular, our goal is to describe the fundamental properties of systems consisting of more than two identical particles. Secondly, the indistinguishability was recognized  as a resource for quantum computation \cite{aaronson2011computational}, therefore its deeper understanding can result in future practical applications. 


\emph{General framework.--} Every experiment has three stages. The first stage is a preparation of a system in some initial state. Due to various reasons the state needs not to be exactly determined. Therefore, the most general state description is given by a set of probability distributions over the values of measurable properties. In the next stage the system undergoes an evolution and its state changes. The description of this change is given by a set of allowable transformations on the set of probability distributions. Finally, in the last stage some properties of the system are measured.

In this work we consider a system of $N$ noninteracting identical particles which can be distributed over $K$ different modes. The state is determined by a set of particle occupation numbers for each mode $s=\{n_1,n_2,\ldots,n_K\}$. The number of particles is conserved, therefore $\sum_{i=1}^K n_i=N$. The total number of different states is $d=\frac{(K+N-1)!}{N!(K-1)!}$.  Since the description needs not to be deterministic, we consider d-dimensional probability vectors $\mathbf{\Pi}$ over all states. We will  refer to these vectors as distributions.

The model is simple -- we prepare an initial probability distribution $\mathbf{\Pi_{i}}$ which is transformed into a final distribution $\mathbf{\Pi_{f}}$. The transformation is given by a stochastic matrix $\mathbb{S}$, i.e., $\mathbf{\Pi_{f}}=\mathbb{S}\mathbf{\Pi_{i}}$. The distribution $\mathbf{\Pi_{f}}$ describes the statistics of detection events which can be registered by particle counters.

In order to illustrate the above idea let us consider a well known example of $N=2$ and $K=2$ corresponding to noninteracting bosons on a symmetric beam-splitter (BS). There are three possible states, which we denote as $\{2,0\},\,\{1,1\}$ and $\{0,2\}$. The probability vector is of the form $\mathbf{\Pi}=\left(p(2,0),\,p(1,1),\,p(0,2)\right)^T$. The transformation reads
\begin{equation}
\mathbb{S}_{BS}=\begin{pmatrix} 1/4 & 1/2 & 1/4 \\ 1/2 & 0 & 1/2 \\1/4 & 1/2 & 1/4 \end{pmatrix}.
\end{equation}  

Before we proceed, we need to make one important comment. One may question that the above approach does not allow to describe transformations on all physically accessible initial states. For example, our generalized probabilistic framework does not consider quantum superpositions of states $\{2,0\}$, $\{1,1\}$ and $\{0,2\}$. The model assumes that we only deal with mixtures over states with well defined occupation numbers. However, note that any quantum superposition can be obtained from such states by a proper transformation. This pre-transformation can be included in the main transformation. For example, before the particles go into BS, they can go through another device which will prepare a superposition. In a similar way one may question that we do not allow to measure all states. However, just like with preparation, any measurement basis can be transformed into the occupation number basis and this post-transformation can be also included in the main transformation. 


\emph{Consistency condition.--} Let us focus on how to encode the lack of interaction into our framework. We start with an observation regarding a BS transformation made by two of the authors previously in \cite{bosoncontext}. Namely, the transformation of a single-particle distribution does not depend on the presence of the other particle. Here, we generalize this property to arbitrary transformations and arbitrary subsets of particles.

In order to do that we investigate the relationship between the N and (N-1)-partite probability distributions   $\mathbf{\Pi}^{(N)}$ and  $\mathbf{\Pi}^{(N-1)}$. In essence, $\mathbf{\Pi}^{(N-1)}$ should be consistent with a probability distribution obtained from $\mathbf{\Pi}^{(N)}$ by randomly removing one particle. This can be described as  $\mathbf{\Pi}^{{(N-1)}}=\mathbb{D}^{(N)}\mathbf{\Pi}^{{(N)}}$, where  $\mathbb{D}^{(N)}$ is a rectangular stochastic matrix. Its entries $D_{ij}$ correspond to probabilities of transition between an N-partite state  $s_j$ and an (N-1)-partite state $s'_i$. Such a transition  is possible iff deleting a single particle from some mode k of a state $s_j$ gives a state $s'_i$. Let $n_{k(i,j)}$ be the occupation number of the mode that we delete a particle from in $s_j$ to achieve the transition. Then $D_{ij} = n_{k(i,j)}/N$ if this state transition is possible and 0 otherwise.

For example, in the case of $N=2$ and $K=2$ the transition from a bipartite distribution to a single-partite distribution (supported on states $\{1,0\}$ and $\{0,1\}$) is given by the $2 \times 3$ matrix 
\begin{equation}\label{BS2}
\mathbb{D}^{(2)}=\frac{1}{2} \begin{pmatrix} 2 & 1 & 0 \\ 0 & 1 & 2 \end{pmatrix}.
\end{equation}
Finally, one can construct matrices allowing to transform N-partite into M-partite distributions via simple multiplication $\mathbb{D}^{(N\rightarrow M)} = \mathbb{D}^{(M+1)}\ldots \mathbb{D}^{(N-1)}\mathbb{D}^{(N)}$.

Now we introduce constraints on transformations $\mathbb{S}$. We start with a transformation of a single particle $\mathbb{S}^{(1)}$. This transformation is the primitive of our model since due to no interactions single-particle transformation must be the basis for the evolution of an arbitrary number of particles. For example, in the case of a symmetric BS this transformation is given by
\begin{equation}\label{BSt}
\mathbb{S}^{(1)}_{BS}=\frac{1}{2}\begin{pmatrix} 1 & 1 \\ 1 & 1 \end{pmatrix}.
\end{equation}
The bipartite transformation $\mathbb{S}^{(2)}$ can be chosen in an arbitrary way, provided that the following constraint is fulfilled for all bipartite probability vectors $\mathbf{ \Pi_i}^{{(2)}}$:
\begin{equation}\label{principle2}
\mathbb{D}^{(2)} \mathbb{ S}^{(2)} \mathbf{ \Pi_i}^{{(2)}} = \mathbb{S}^{(1)} \mathbb{D}^{(2)} \mathbf{ \Pi_i}^{{(2)}}.
\end{equation}
In simple words, the above means that if we first transform a bipartite distribution and then reduce it to a single-partite distribution we would get the same result as if we first reduced a bipartite distribution to a single-partite distribution and then transformed it. This can be easily generalized to an arbitrary number of particles
\begin{equation}\label{principle}
\mathbb{D}^{(N\rightarrow 1)} \mathbb{S}^{(N)} \mathbf{\Pi_i}^{{(N)}} = \mathbb{S}^{(1)} \mathbb{D}^{(N\rightarrow 1)} \mathbf{ \Pi_i}^{(N)}.
\end{equation}
Moreover, this constraint should hold at the level of all the M-partite subsets 
\begin{equation}\label{NIprinciple}
\mathbb{D}^{(N\rightarrow M)}\mathbb{S}^{(N)} \mathbf{\Pi_i}^{(N)} = \mathbb{S}^{(M)} \mathbb{D}^{(N\rightarrow M)} \mathbf{ \Pi_i}^{(N)},
\end{equation}
where $M$ is an arbitrary integer $M<N$. We will call the equation (\ref{NIprinciple}) the \emph{consistency condition}.

To illustrate this restriction let us once more consider the example of a symmetric BS. Equations (\ref{BS2}), (\ref{BSt}) and (\ref{principle2}) imply $\mathbb{D}^{(2)} \mathbb{S}^{(2)} \mathbf{\Pi_i}^{(2)}=(1/2,  1/2 )^T$ for any distribution $\mathbf{\Pi_i}^{{(2)}}$. This means that the final distribution must satisfy 
\begin{equation}
\begin{pmatrix} p(2,0) + 1/2\, p(1,1) \\ p(0,2) + 1/2\, p(1,1) \end{pmatrix} = \begin{pmatrix} 1/2 \\ 1/2 \end{pmatrix}.
\end{equation}
Note, that the above is obeyed by bosons, fermions and distinguishable particles as well. 

\emph{No-erasure of information.--} Additionally we would like our model to capture the fact that quantum multiports do not erase information. This property can be easily encoded in the transformation matrix $\mathbb{S}$ by requiring it to be doubly stochastic, i.e. the sum of its entries in each row and column equals to one \cite{cover}. 

For instance, in the case of an asymmetric beam splitter the matrix $\mathbb{S}$ is given by
\begin{equation}\label{aBSt}
\mathbb{S}^{(1)}_{aBS}=\begin{pmatrix} T & R \\ R & T \end{pmatrix},
\end{equation}
where $T+R=1$.


\emph{Beyond quantum theory.--} Multipartite quantum states are expressed in terms of operators $a^{\dagger}_i$ which create a particle in mode $i$. For our purposes we do not need to go into details about the underlying particle-statistics to show that the consistency condition is obeyed in quantum theory. It is enough to observe that due to lack of interaction creation operators evolve independently $a^{\dagger}_i \rightarrow a'^{\dagger}_i$. Therefore, (\ref{NIprinciple}) is automatically satisfied. Moreover, since the evolution operator is unitary, the entropy of the system does not decrease. It might increase due to the final measurement, but it never goes down. Because of that, the no-erasure condition is also observed.

Interestingly, the two requirements still allow for a more general description of transformations. Although the quantum theory admits perfect bunching and anti-bunching in the $N=2$ and $K=2$ scenario, for $N>2$ or $K>2$ one can propose some more extreme behaviors. Here, we discuss the case $N=3$ and $K=3$.

Let us first consider a quantum description of a symmetric three-port, commonly known as a tritter. Its quantum properties have been studied in great details -- see for example the work by Campos \cite{PhysRevA.62.013809}. There are three input modes described by creation operators $a_i^{\dagger}$ and three output modes described by $a'^{\dagger}_i$ ($i=1,2,3$). The transformation is given by a unitary mapping $a'^{\dagger}_i =  \sum_j U_{ij} a^{\dagger}_j$, where $ U_{ij}=\frac{1}{\sqrt 3} \omega^{\delta_{ij}}$, $\delta_{ij}$ is the Kronecker delta and $\omega$ is the third root of unity. 

The quantum tritter transformation applied to the three-boson state $\{1,1,1\}$ produces states $\{3,0,0\}$, $\{0,3,0\}$ and $\{0,0,3\}$ with a probability $2/9$ each and state $\{1,1,1\}$ with a probability $1/3$. Interestingly, unlike $N=2$ and $K=2$ the quantum probability of bunching $B_Q$ does not saturate the algebraic bound of one 
\begin{equation}
B_Q=p_{300}^{(111)}+p_{030}^{(111)}+p_{003}^{(111)}=\frac{2}{3}<1,
\end{equation}
where $P_x^{(y)}$ denotes the probability of transforming the state y into state x. Although we do not provide a proof of this statement, by the end of this work we will show that the value of $2/3$ is implied by a fundamental principle obeyed by the quantum theory. 

At this point one may wonder if some hypothetical particles, which obey the consistency and no-erasure conditions, can have greater tripartite-bunching properties than bosons.  The answer is positive. Consider for example the three-particle transformation $\mathbb{S}^{(3)}_T$ 

\begin{eqnarray}
&&\begin{array}{ccccccccccc}
{ \:  \,\,
 \color{white} 000000}& 3 & 0 & 0\; &\; 2 & 2 & 1 & 1& 0 & 0\; &\;\;\;\: 1 \\
{\color{white} 000000}& 0 & 3 & 0\; &\; 1 & 0 & 2 & 0& 2 & 1\; &\; \;\;\:1\\
{\color{white} 000000}& 0 & 0 & 3\; &\; 0 & 1 & 0 & 2& 1 & 2\; &\; \;\;\:1 \\
\end{array}\nonumber\\
&&\begin{array}{ccc}
3 & 0 & 0\\
0 & 3 & 0\\
0 & 0 & 3\\
&\\

2 & 1 & 0\\
2 & 0 & 1\\
1 & 2 & 0\\
1 & 0 & 2\\
0 & 2 & 1\\
0 & 1 & 2\\
&  \\
1 & 1 & 1\\
&\\

\end{array}
\left[
\begin{array}{c|c|c}
 &  \\
~~~0~~~\;\; & ~~~~~~\;\;\,\dfrac{1}{8}\;\,\;~~~~~~ & ~~\dfrac{1}{4}~~ \\
&  \\
\hline
 &  \\
 &\\
 &  \\
\dfrac{1}{8} & \dfrac{5}{48} & ~~0~~\\
&  \\
&  \\
\hline
 &  \\
\dfrac{1}{4} & 0 & ~~\dfrac{1}{4}~~\\
&  \\

\end{array}
\right],
\label{3tritter}
\end{eqnarray} 
where the blocks are schematically denoted by a single element which is the same for all its entries. The above transformation has a bunching probability $B_S=\frac{3}{4}$. One can easily verify that  $\mathbb{S}^{(3)}_T$  is doubly stochastic. It remains to be shown that it also follows the consistency condition. In order to do that, note that from the point of view of our generalized probabilistic description the single-partite tritter transformation is given by
\begin{equation}
\mathbb{S}^{(1)}_{T} = \frac{1}{3} \begin{pmatrix} 1 &1 & 1 \\ 1 &1 & 1 \\ 1 &1 & 1 \end{pmatrix}.
\end{equation}
Then for all the possible initial states we indeed have  $\mathbb{D}^{(3\rightarrow 1)} \mathbb{S}_T^{(3)} \mathbf{ \Pi_i}^{(3)}=\mathbb{S}^{(1)}_T \mathbb{D}^{(3\rightarrow 1)} \mathbf{ \Pi_i}^{(3)}$.


\emph{Recovering quantum theory.--} The above example can be considered as an identical-particle analog of the PR-box \cite{Popescu1994}. Note, that in the case of the PR-boxes the no-signalling principle does not recover quantum theory. In our case the conditions imposed by our framework do not recover quantum theory either. However, here we find an additional physical restriction that allows for recovery of bosonic behaviour on a tritter.

Let us first observe that the average two-particle bunching probability bounds the three-particle one from above.
\begin{equation}
B\leq \frac{p_{200}^{(110)}+p_{200}^{(101)}+p_{200}^{(011)}+...+p_{002}^{(101)}+p_{002}^{(011)}}{3}.
\end{equation}
 This follows form the consistency condition applied to the general form of the transformation of state $\{1,\,1,\,1\}$.
Since the transformation matrix needs to be doubly stochastic the above expression can be written as
\begin{equation}\label{2bunch}
B\leq 1- \frac{p_{200}^{(200)}+p_{200}^{(020)}+p_{200}^{(002)}+...+p_{002}^{(020)}+p_{002}^{(002)}}{3}.
\end{equation}
The goal is therefore to show that the right hand side of (\ref{2bunch}) is bounded from above by $\frac{2}{3}$. To do that, we propose a restriction on the possible transformations of a certain class of states.

Let us introduce it in the case of two particles. We say that a 2-particle system corresponding to a probability distribution $\mathbf{\Pi}^{(2)}$ is composite iff it satisfies the entropic relation 
\begin{equation}
H(\mathbf{\Pi}^{(2)})=2H(\mathbf{\Pi}^{(1)})=2H(\mathbb D^{(2\rightarrow 1)}\mathbf \Pi^{(2)}).
\end{equation}
 This condition states that the information content of  the whole system is the same as the sum of information contents of its two one-particle subsystems. It is easy to verify that the only composite system of two particles corresponds to a state of the form $\{2,\,0,\,0\}$. As a side note observe that in the first quantization these are the only states of identical particles that can be written in a product form. Consequences of that fact have been studied in \cite{Loudon,Bach}.

Now we propose a  principle which states that a composite system evolves as a product of  evolutions of its subsystems. In the 2-particle case this means that a composite  system $\mathbf \Pi^{(2)}$ evolves as
\begin{equation}\label{product}
\mathbb S^{(2)}\mathbf \Pi^{(2)}=\mathbb S^{(1)}\mathbf \Pi^{(1)}\times \mathbb S^{(1)}\mathbf \Pi^{(1)},
\end{equation}
where states of the type $a\times b$ are treated as equivalent to $b \times a$ because of the indistinguishability of particles. For instance we have
\begin{equation}
\begin{split}
\{0,\,1,\,1\}&\equiv\{0,\,1,\,0\}\times\{0,\,0,\,1\}\\
&\equiv\{0,\,0,\,1\}\times\{0,\,1,\,0\},
\end{split}
\end{equation}
so the two-particle vector space shrinks to 6 dimensions.  Formula (\ref{product}) means that for a tritter we have $p_{200}^{(200)}=p_{020}^{(200)}=p_{002}^{(200)}=\frac{1}{9}$. Since the same reasoning holds for the states $\{0,\,2,\,0\}$ and $\{0,\,0,\,2\}$, inequality (\ref{2bunch}) simplifies to
\begin{equation}
B\leq\frac{2}{3},
\end{equation}
which is the quantum bound for a three-partite bunching on a tritter.

Finally, we would like to note that our approach can be applied to any system, not only the tritter. For instance, two particles on an N-port have the average quantum bunching probability $B$ equal to
\begin{equation}
B=2\,\frac{p_{20\cdots0}^{(11\cdots0)}+\cdots+p_{0\cdots02}^{(0\cdots011)}}{N(N-1)}=\frac{2}{N},
\end{equation}
which is also the upper bound on bunching in our model. On the other hand, some additional restrictions are required to recover quantum behavior for $N>3$ and $K>3$. For example, we have considered the case $N=4$ and $K=4$ and observed that the
following transformation admits superquantum bunching while satisfying all the restrictions of our model

\begin{eqnarray}
&&\begin{array}{cccccccccc}
{\,\,\,\,\, \:  \,\,
 \color{white} 000000}& 3 & \cdots & 0\; &\; 2 & \cdots & 0 \;&\; 1& \cdots & 0\; \\
{\color{white} 000000}& 0 & \cdots & 0\; &\; 1 & \cdots & 0 \;&\; 1& \cdots & 1\;\\
{\color{white} 000000}& 0 & \cdots & 0\; &\; 0 & \cdots & 1 \;&\; 1& \cdots & 1\; \\
{\color{white} 000000}& 0 & \cdots & 3\; &\; 0 & \cdots & 2 \;&\; 0& \cdots & 1\; \\
\end{array}\nonumber\\
&&\begin{array}{cccc}
3 & 0 & 0 &0\\
\vdots&\vdots & \vdots&\vdots\\
0 & 0 & 0&3\\
&\\

2 & 1 & 0&0\\
\vdots&\vdots & \vdots&\vdots\\
0 & 0 & 1&2\\
&\\
1 & 1 & 1&0\\
\vdots&\vdots & \vdots&\vdots\\
0 & 1 & 1 &1\\

\end{array}
\left[
\begin{array}{c|c|c}
~~~\dfrac{3}{32}~~~ &\,\,~~~\dfrac{3}{64}~~~ \,\,&~~~\dfrac{1}{64}~~~  \rule{0pt}{6ex} \rule[-6ex]{0pt}{0pt} \\

\hline
 
\dfrac{1}{96} &\dfrac{13}{192}&\dfrac{7}{192} \rule{0pt}{6.7ex} \rule[-6.0ex]{0pt}{0pt} \\

\hline
\dfrac{1}{8} &0 &\dfrac{1}{8} \rule{0pt}{6.5ex} \rule[-5.0ex]{0pt}{0pt} \\


\end{array}
\right]
\label{3multi}
\end{eqnarray} 
Perhaps an extended version of the composite system argument is needed to explain all but the simplest cases.

\emph{ Conclusions and outlook.--} We have proposed an operational description of the evolution of non-interacting indistinguishable particles. The model is particularly related to linear optical experiments with multipartite interference of bosons or fermions. Our approach explores bunching in generalized theories. In particular, we show that our framework admits exotic bunching probabilities. The superquantum symmetric 3-port (tritter) we present could be considered a PR-box counterpart in the realm of particle statistics. In this case an additional principle, governing the evolution of a certain class of states, is enough to recover the bosonic bounds. However, it is not sufficient in higher-order multiports. There are few possibilities why this happens. Firstly, our principle may need extension to include more general classes of states, not only those in which all the particles are in the same mode. Moreover, symmetric quantum n-ports (for $n>3$) have more than one inequivalent representation \cite{mattle}. For example, a 4-port can be described by a discrete Fourier transform, but also by a Groover-like unitary matrix. The two representations generate different output probability distributions. Nevertheless, they  both lead to the same maximal bunching probabilities, which suggest that fundamental laws behind bosonic behavior go beyond the subtleties of particular evolutions (see the Appendix for details).

This work fits into the very lively field of  research on explaining elements of the quantum mechanics with intuitive principles.   Investigation of  particle statistics admitted by superquantum theories is of general interest, as it might lead to new tests of  quantum foundations. We hope that our results will be a stimulating contribution to this endeavor.


\emph{Acknowledgements.--} We would like to thank Micha{\l} Oszmaniec and Andrzej Grudka for valuable comments. M.K., M.M. and P.K. were supported by the National Science Centre in Poland through NCN Grant No. 2014/14/E/ST2/00585. D.K. was supported by the National Research Foundation and Ministry of Education in Singapore.



\section{Appendix}

In general, a quantum n-port is described by an operator $a'^{\dagger}_j =  \sum_k U_{jk} a^{\dagger}_k$, where creation operators $a_j^{\dagger}$ and $a'^{\dagger}_j$ correspond to $j^{th}$ input and output modes and $U$ is a unitary matrix of coefficients. Such an n-port is called symmetric if a single particle cast on any input mode is transformed into every output mode with equal probability. This condition is satisfied when
 \begin{equation}\label{sym}
 |U_{jk}|=\frac{1}{\sqrt{n}}
 \end{equation} for every $j$ and $k$.

Given the transformation of creation operators one can easily calculate the corresponding probabilities of state transformations. For a symmetric 3-port a 3-particle matrix of transformation probabilities is given by 
\begin{eqnarray}
&&\begin{array}{cccccccc}
{\,\,\,\,
 \color{white} 000000}& 3 & \cdots & 0\; &\; 2 & \cdots & 0 \;&\;\;~~~ 1\;\; \\
{\color{white} 000000}& 0 & \cdots & 0\; &\; 1 & \cdots & 1 \;&\;~~~ 1 \;\\
{\color{white} 000000}& 0 & \cdots & 3\; &\; 0 & \cdots & 2 \;&\;~~~ 1\; \\

\end{array}\nonumber\\
&&\begin{array}{ccc}
3 & 0 & 0 \\
\vdots&\vdots & \vdots\\
0 & 0 & 3\\
&\\

2 & 1 & 0\\
\vdots&\vdots & \vdots\\
0 & 1 & 2\\
&\\
1 & 1 & 1\\


\end{array}
\left[
\begin{array}{c|c|c}
~~~~\dfrac{1}{27}~~~~ &\,\,~~~~\dfrac{1}{9}~~~~ \,\,&~~~~\dfrac{2}{9}~~~~  \rule{0pt}{6ex} \rule[-6ex]{0pt}{0pt} \\

\hline
 
\dfrac{1}{9} &\dfrac{1}{9}&0 \rule{0pt}{7ex} \rule[-6ex]{0pt}{0pt} \\

\hline
\dfrac{2}{9} &0 &\dfrac{1}{3} \rule{0pt}{5ex} \rule[-1.0ex]{0pt}{0pt} \\


\end{array}
\right].
\label{3multi}
\end{eqnarray} \\
Interestingly, these 3-port probabilities do not depend on the choice of $U$ provided that (\ref{sym}) is observed. Moreover, the matrix consists of blocks that guarantee that the transformation probabilities do not change when input or output modes are permuted. These two properties do not hold for 4-ports.

To see that one can consider the coefficients matrices of form $ U^{(F)}_{jk}=\frac{1}{2} i^{(j-1)(k-1)}$ and $ U^{(G)}_{jk}=\frac{1}{2} (1-2\delta_{j,\,k})$ which correspond to Fourier and  Grover-like 4-ports. In the simplest nontrivial case of two particles these coefficients lead to the following transformation probabilities.

\begin{eqnarray}
&&\begin{array}{cccccccccccc}
{\,\,\,\,\,\,\,\,\;\;\;
 \color{white} 000000}& 2 & \cdots & 0\; & 1 \;&\;1\;&\;1\;&\;0\;&\;0\;&\;0\; \\
{\color{white} 000000}& 0 & \cdots & 0\; & 1 \;&\;0\;&\;0\;&\;1\;&\;1\;&\;0\;\\
{\color{white} 000000}& 0 & \cdots & 0\; & 0 \;&\;1\;&\;0\;&\;1\;&\;0\;&\;1\; \\
{\color{white} 000000}& 0 & \cdots & 2\; & 0 \;&\;0\; &\;1\;&\;0\;&\;1\;&\;1\;\\

\end{array}\nonumber\\
&&\begin{array}{cccc}
2 & 0 & 0&0 \\
\vdots&\vdots & \vdots& \vdots\\
0 & 0 & 0& 2\\
&\\

1 & 1 & 0&0\\
1 & 0 & 1&0\\
1 & 0 & 0&1\\
0 & 1 & 1&0\\
0 & 1 & 0&1\\
0 & 0 & 1&1\\


\end{array}
\left[
\begin{array}{c|c}
~~\;~\dfrac{1}{16}\;~~~ &\,\,~~~~~~~~~~\dfrac{1}{8}~~~~~~~~~~ \,\,  \rule{0pt}{6ex} \rule[-6ex]{0pt}{0pt} \\

\hline
 
\dfrac{1}{8} & \scalebox{2}A \rule{0pt}{11ex} \rule[-9ex]{0pt}{0pt} \\




\end{array}
\right],
\label{3multi}
\end{eqnarray} 

where $A$ denotes a $6\times 6$ block of form

\begin{eqnarray}
& &\left[
\begin{array}{cccccc}
\;\;\dfrac{1}{8}\;\;&\;\;0&\dfrac{1}{8}\;\;&\;\;\dfrac{1}{8}\;\;&\;\;0\;\;&\;\;\dfrac{1}{8}\;\;\\\\
0&\dfrac{1}{4}&0&0&\dfrac{1}{4}&0\\\\
\;\dfrac{1}{8}\;&\;0&\dfrac{1}{8}\;&\;\dfrac{1}{8}\;&\;0\;&\;\dfrac{1}{8}\;\\\\
\;\dfrac{1}{8}\;&\;0&\dfrac{1}{8}\;&\;\dfrac{1}{8}\;&\;0\;&\;\dfrac{1}{8}\;\\\\
0&\dfrac{1}{4}&0&0&\dfrac{1}{4}&0\\\\
\;\dfrac{1}{8}\;&\;0&\dfrac{1}{8}\;&\;\dfrac{1}{8}\;&\;0\;&\;\dfrac{1}{8}\;
\end{array}
\right]
\end{eqnarray}
for the Fourier multiport and
\begin{eqnarray}
& &\left[
\begin{array}{cccccc}
\;\;\dfrac{1}{4}\;\;&\;\;0\;\;&\;\;0\;\;&\;\;0\;\;&\;\;0\;\;&\;\;\dfrac{1}{4}\;\;\\\\
0&\dfrac{1}{4}&0&0&\dfrac{1}{4}&0\\\\
0&0&\dfrac{1}{4}&\dfrac{1}{4}&0&0\\\\
0&0&\dfrac{1}{4}&\dfrac{1}{4}&0&0\\\\
0&\dfrac{1}{4}&0&0&\dfrac{1}{4}&0\\\\

\dfrac{1}{4}&0&0&0&0&\dfrac{1}{4}
\end{array}
\right]
\end{eqnarray}
for the Grover-like one. Although these matrices differ, they are both consistent with the framework and additional principle proposed in our paper.

\end{document}